\documentclass[10pt,conference,letterpaper,final]{IEEEtran}

\usepackage[T1]{fontenc}
\usepackage[utf8]{inputenc}

\usepackage{algorithmic}
\usepackage{amsmath}
\usepackage{array}
\usepackage[pdftex]{graphicx}
\usepackage{placeins}
\usepackage{todonotes}
\usepackage{url}
\usepackage{xspace}
\usepackage{tikz}
\usetikzlibrary{bending,positioning,calc,patterns,backgrounds}

\usepackage{fixltx2e}
\usepackage{dblfloatfix}
\usepackage[hidelinks]{hyperref}

\usepackage[noadjust]{cite}

\usepackage{xfrac}
\usepackage[detect-all]{siunitx}
\begin{document}

% renewed commands

\renewcommand{\rm}{{r_\mathrm{m}}\xspace}
\renewcommand{\and}{{\mathrm{and}}\xspace}
\newcommand{\paragraphlb}[1]{\paragraph{#1}\mbox{}\\\vspace{-10pt}}

% command shorthands

\newcommand{\bs}[1]{\boldsymbol{#1}}
\newcommand{\minisection}[1]{\paragraph{\emph{#1}}}
\newcommand{\paramtype}[2]{\mathsf{#1_{#2}}}
\newcommand{\sidenote}[1]{\marginnote{\emph{#1}}}
\newcommand{\tb}[1]{\textbf{#1}}
\newcommand{\ti}[1]{\textit{#1}}
\newcommand{\todoin}[1]{\todo[inline]{#1}}
\newcommand{\todosolved}[1]{\todo[inline, color=green]{#1}}

% math shorthands

\newcommand{\Bigcap}{\bigcap\limits}
\newcommand{\DKL}[2]{D_\mathrm{KL}\left(#1\parallel#2\right)}
\newcommand{\DKLnorm}[2]{D_\mathrm{KL}^\mathrm{norm}\left(#1\parallel#2\right)}
\newcommand{\Expect}[1]{E\left[#1\right]}
\newcommand{\expect}[1]{\left\langle#1\right\rangle}
\newcommand{\Int}{\int\limits}
\newcommand{\Lim}{\lim\limits}
\newcommand{\Prod}{\prod\limits}
\newcommand{\Sum}{\sum\limits}
\newcommand{\var}[1]{\Var\left[#1\right]}
\newcommand\asymunc[4]{\ensuremath{#1\,\substack{+#2\\-#3}}\,#4}

% new symbols

\newcommand{\ci}[3]{#1 {\perp\!\!\!\perp} #2 \; | \; #3}
\newcommand{\nci}[3]{#1 \centernot{\perp\!\!\!\perp} #2 \; | \; #3}

% indices with \mathrm stuff

\newcommand{\abstr}{\mathrm{abstr}\xspace}
\newcommand{\bmErev}{\bm{E}^\mathrm{rev}\xspace}
\newcommand{\bmtausyn}{\bm\tau^\mathrm{syn}\xspace}
\newcommand{\bmueff}{\bm{u}_\mathrm{eff}\xspace}
\newcommand{\boxfct}{\mathrm{box}\xspace}
\newcommand{\Ca}{{\mathrm{Ca}^{++}}\xspace}
\newcommand{\CC}{\rho\xspace}
\newcommand{\Cl}{{\mathrm{Cl}^-}\xspace}
% collides with siunitx
% \newcommand{\cm}{{c_\mathrm{m}}\xspace}
\newcommand{\Cm}{{C_\mathrm{m}}\xspace}
\newcommand{\const}{{\mathrm{const}}\xspace}
\newcommand{\Cref}{{C_\mathrm{ref}}\xspace}
\newcommand{\CVISI}{{\mathrm{CV}_\mathrm{ISI}}\xspace}
\newcommand{\CVrate}{{\mathrm{CV}_\mathrm{rate}}\xspace}
\newcommand{\ddt}{{\frac{d}{dt}}\xspace}
\newcommand{\DeltaT}{{\Delta_\mathrm{T}}\xspace}
\newcommand{\DKLsolo}{{D_\mathrm{KL}}\xspace}
\newcommand{\El}{{E_\mathrm{l}}\xspace}
\newcommand{\Eqn}{\mathrm{Eqn.}\xspace}
\newcommand{\Eqns}{\mathrm{Eqns.}\xspace}
\newcommand{\Er}{{E_\mathrm{r}}\xspace}
\newcommand{\Erev}{E^\mathrm{rev}\xspace}
\newcommand{\Ereve}{{E^\mathrm{rev}_\mathrm{e}}\xspace}
\newcommand{\Erevi}{{E^\mathrm{rev}_\mathrm{i}}\xspace}
\newcommand{\ET}{{E_\mathrm{T}}\xspace}
\newcommand{\gext}{{g_\mathrm{ext}}\xspace}
\newcommand{\gimax}{{g_i^\mathrm{max}}\xspace}
\newcommand{\gl}{{g_\mathrm{l}}\xspace}
\newcommand{\fsyn}{{f^\mathrm{syn}}\xspace}
\newcommand{\gsyn}{g^\mathrm{syn}\xspace}
\newcommand{\gsyne}{{g^\mathrm{syn}_\mathrm{e}}\xspace}
\newcommand{\gsyni}{{g^\mathrm{syn}_\mathrm{i}}\xspace}
\newcommand{\gtot}{{g^\mathrm{tot}}\xspace}
\newcommand{\icb}{\texttt{icb}}
\newcommand{\iext}{{i^\mathrm{ext}}\xspace}
\newcommand{\Iext}{I^\mathrm{ext}\xspace}
\newcommand{\ifmath}{{\mathrm{if}}\xspace}
\newcommand{\Inoise}{I^\mathrm{noise}}
\newcommand{\ipi}{{}^1p_1\xspace}
\newcommand{\irc}{{i^\mathrm{RC}}\xspace}
\newcommand{\Irec}{I^\mathrm{rec}}
\newcommand{\Iref}{{I_\mathrm{ref}}\xspace}
\newcommand{\isyn}{i^\mathrm{syn}\xspace}
\newcommand{\Isyn}{I^\mathrm{syn}\xspace}
\newcommand{\Jsyn}{J^\mathrm{syn}\xspace}
% collides with siunitx
% \newcommand{\K}{{\mathrm{K}^+}\xspace}
\newcommand{\lambdam}{{\lambda_\mathrm{m}}\xspace}
\newcommand{\Mexc}{M_\mathrm{exc}\xspace}
\newcommand{\Minh}{M_\mathrm{inh}\xspace}
\newcommand{\Na}{{\mathrm{Na}^+}\xspace}
\newcommand{\NBAS}{{N_\mathrm{BAS}}\xspace}
\newcommand{\nne}{{\mathrm{ne}}\xspace}
\newcommand{\NHC}{{N_\mathrm{HC}}\xspace}
\newcommand{\NMC}{{N_\mathrm{MC}}\xspace}
\newcommand{\non}{{\mathrm{\setminus}}\xspace}
\newcommand{\nonk}{{\non k}\xspace}
\newcommand{\NPYR}{{N_\mathrm{PYR}}\xspace}
\newcommand{\nRS}{{n_\mathrm{RS}}\xspace}
\newcommand{\nFS}{{n_\mathrm{FS}}\xspace}
\newcommand{\nusyn}{\nu^\mathrm{syn}\xspace}
\newcommand{\NRSNP}{{N_\mathrm{RSNP}}\xspace}
\newcommand{\otherwise}{\mathrm{otherwise}\xspace}
\newcommand{\pa}{\mathrm{\textbf{pa}}\xspace}
\newcommand{\pflip}{p_\mathrm{flip}\xspace}
\newcommand{\poo}{p_{00}\xspace}
\newcommand{\poi}{p_{01}\xspace}
\newcommand{\pio}{p_{10}\xspace}
\newcommand{\pii}{p_{11}\xspace}
\newcommand{\PSP}{\mathrm{PSP}\xspace}
\newcommand{\pspike}{p_\mathrm{spike}\xspace}
\newcommand{\rl}{{r_\mathrm{l}}\xspace}
\newcommand{\Rtest}{{R^\mathrm{test}}\xspace}
\newcommand{\Rtrain}{{R^\mathrm{train}}\xspace}
\newcommand{\SU}{\tilde I\xspace}
\newcommand{\taubk}{\overline{\tau^\mathrm{b}_k}}
\newcommand{\taudecay}{{\tau_\mathrm{decay}}\xspace}
\newcommand{\taueff}{{\tau_\mathrm{eff}}\xspace}
\newcommand{\taufacil}{{\tau_\mathrm{facil}}\xspace}
\newcommand{\taufall}{{\tau_\mathrm{fall}}\xspace}
\newcommand{\tauinact}{{\tau_\mathrm{inact}}\xspace}
\newcommand{\taum}{{\tau_\mathrm{m}}\xspace}
\newcommand{\tauon}{{\tau_\mathrm{on}}\xspace}
\newcommand{\tauON}{{\tau_\mathrm{ON}}\xspace}
\newcommand{\taurise}{{\tau_\mathrm{rise}}\xspace}
\newcommand{\taurec}{{\tau_\mathrm{rec}}\xspace}
\newcommand{\tauref}{{\tau_\mathrm{ref}}\xspace}
\newcommand{\taustdp}{\tau^\mathrm{STDP}\xspace}
\newcommand{\tausyn}{\tau^\mathrm{syn}\xspace}
\newcommand{\tausyne}{{\tau^\mathrm{syn}_\mathrm{e}}\xspace}
\newcommand{\tausyni}{{\tau^\mathrm{syn}_\mathrm{i}}\xspace}
\newcommand{\tauw}{{\tau_\mathrm{w}}\xspace}
\newcommand{\textmax}{{\mathrm{max}}\xspace}
\newcommand{\textmin}{{\mathrm{min}}\xspace}
\newcommand{\thetaeff}{{\vartheta_\mathrm{eff}}\xspace}
\newcommand{\trise}{{t_\mathrm{rise}}\xspace}
\newcommand{\tfall}{{t_\mathrm{fall}}\xspace}
\newcommand{\tspike}{{t_\mathrm{spike}}\xspace}
\newcommand{\ueff}{{u_\mathrm{eff}}\xspace}
\newcommand{\unif}{{\mathrm{unif}}\xspace}
\newcommand{\ureset}{{u_\mathrm{reset}}\xspace}
\newcommand{\USE}{{U_\mathrm{SE}}\xspace}
\newcommand{\uthr}{{u_\mathrm{thr}}\xspace}
\newcommand{\Vm}{{V_\mathrm{m}}\xspace}
\newcommand{\Vrest}{{V_\mathrm{rest}}\xspace}
\newcommand{\Vspike}{{V_\mathrm{spike}}\xspace}
\newcommand{\Vth}{{V_\mathrm{th}}\xspace}
\newcommand{\Vthresh}{\ET}
\newcommand{\wsyn}{w^\mathrm{syn}\xspace}
\newcommand{\zpi}{{}^2p_1\xspace}

% text in equations stuff

\newcommand{\COBA}{\mathrm{COBA}\xspace}
\newcommand{\Cov}{\mathrm{Cov}\xspace}
\newcommand{\CUBA}{\mathrm{CUBA}\xspace}
\newcommand{\erf}{\mathrm{erf}\xspace}
\newcommand{\for}{\mathrm{for}\xspace}
\newcommand{\sgn}{\mathrm{sgn}\xspace}
\newcommand{\Var}{\mathrm{Var}\xspace}

% code stuff

\newcommand{\aEIFcurrexp}{\mbox{\texttt{aEIF\_curr\_exp}}\xspace}
\newcommand{\IFcondalpha}{\mbox{\texttt{IF\_cond\_alpha}}\xspace}
\newcommand{\NEST}{\mbox{\texttt{NEST}}\xspace}
\newcommand{\Neuron}{\mbox{\texttt{Neuron}}\xspace}
\newcommand{\PyNN}{\mbox{\texttt{PyNN}}\xspace}

% environment stuff

\newcommand{\tagarray}{\mbox{}\refstepcounter{equation}$(\theequation)$}
\newenvironment{texttab}[1]
    {\vspace{-10pt}
     \tabulinesep=7pt
     \begin{center}
         \begin{tabu} to 1.013\textwidth {#1}}
        {\end{tabu}
    \end{center}}
    
% vitali osumness

\def\layersep{2.5cm} % Gap between visible & hidden units
\def\neuronsep{1.2} % Horizontal gap between neurons
\tikzstyle{neuron}=[circle,fill=black!25,minimum size=21pt,inner sep=0pt]
\tikzstyle{visible neuron}=[neuron, fill=green!50]
\tikzstyle{hidden neuron}=[neuron, fill=orange!75]

\title{Pattern representation and recognition with accelerated analog neuromorphic systems}

\author{%
  \IEEEauthorblockN{\footnotesize
    M.~A.~Petrovici\IEEEauthorrefmark{2}\IEEEauthorrefmark{3}\enspace
    S.~Schmitt\IEEEauthorrefmark{2}\enspace
    J.~Klähn\IEEEauthorrefmark{2}\enspace
    D.~Stöckel\IEEEauthorrefmark{2}\enspace
    A.~Schroeder\IEEEauthorrefmark{2}\enspace
    \\
    G.~Bellec\IEEEauthorrefmark{6}\enspace
    J.~Bill\IEEEauthorrefmark{2}\enspace
    O.~Breitwieser\IEEEauthorrefmark{2}\enspace
    I.~Bytschok\IEEEauthorrefmark{2}\enspace
    A.~Grübl\IEEEauthorrefmark{2}\enspace
    M.~Güttler\IEEEauthorrefmark{2}\enspace
    A.~Hartel\IEEEauthorrefmark{2}\enspace
    S.~Hartmann\IEEEauthorrefmark{4}\enspace
    D.~Husmann\IEEEauthorrefmark{2}\enspace
    K.~Husmann\IEEEauthorrefmark{2}\enspace
    \\
    S.~Jeltsch\IEEEauthorrefmark{2}\enspace
    V.~Karasenko\IEEEauthorrefmark{2}\enspace
    M.~Kleider\IEEEauthorrefmark{2}\enspace
    C.~Koke\IEEEauthorrefmark{2}\enspace
    A.~Kononov\IEEEauthorrefmark{2}\enspace
    C.~Mauch\IEEEauthorrefmark{2}\enspace
    E.~Müller\IEEEauthorrefmark{2}\enspace
    P.~Müller\IEEEauthorrefmark{2}\enspace
    J.~Partzsch\IEEEauthorrefmark{4}\enspace
    T.~Pfeil\IEEEauthorrefmark{2}\enspace
    S.~Schiefer\IEEEauthorrefmark{4}\enspace
    \\
    S.~Scholze\IEEEauthorrefmark{4}\enspace
    A.~Subramoney\IEEEauthorrefmark{6}\enspace
    V.~Thanasoulis\IEEEauthorrefmark{4}\enspace
    B.~Vogginger\IEEEauthorrefmark{4}\enspace
    R.~Legenstein\IEEEauthorrefmark{6}\enspace
    W.~Maass\IEEEauthorrefmark{6}\enspace
    R.~Schüffny\IEEEauthorrefmark{4}\enspace
    C.~Mayr\IEEEauthorrefmark{4}\enspace
    J.~Schemmel\IEEEauthorrefmark{2}\enspace
    K.~Meier\IEEEauthorrefmark{2}
  }
  \IEEEauthorblockA{\footnotesize\IEEEauthorrefmark{2} Heidelberg University, Kirchhoff-Institute for Physics, Im Neuenheimer Feld 227, D-69120 Heidelberg}
  \IEEEauthorblockA{\footnotesize\IEEEauthorrefmark{3} University of Bern, Department of Physiology, Bühlplatz 5, CH-3012 Bern}
  \IEEEauthorblockA{\footnotesize\IEEEauthorrefmark{4}Technische Universität Dresden, Chair of Highly-Parallel VLSI-Systems and Neuromorphic Circuits, D-01062 Dresden}
  \IEEEauthorblockA{\footnotesize\IEEEauthorrefmark{6}Graz University of Technology, Institute for Theoretical Computer Science, A-8010 Graz}
}

\maketitle

\begin{abstract}
    Despite being originally inspired by the central nervous system, artificial neural networks have diverged from their biological archetypes as they have been remodeled to fit particular tasks.
    In this paper, we review several possibilites to reverse map these architectures to biologically more realistic spiking networks with the aim of emulating them on fast, low-power neuromorphic hardware.
    Since many of these devices employ analog components, which cannot be perfectly controlled, finding ways to compensate for the resulting effects represents a key challenge. 
    Here, we discuss three different strategies to address this problem: the addition of auxiliary network components for stabilizing activity, the utilization of inherently robust architectures and a training method for hardware-emulated networks that functions without perfect knowledge of the system's dynamics and parameters.
    For all three scenarios, we corroborate our theoretical considerations with experimental results on accelerated analog neuromorphic platforms.
\end{abstract}

\section{Introduction}

Artificial neural networks (ANNs) rank among the most successful classes of machine learning models, but are -- superficial similarities to sensory processing pathways in cortex notwithstanding -- difficult to map to biologically realistic spiking neural networks.
Nevertheless, we argue that such a reverse mapping is worthwhile for two reasons.
First, it could help us understand information processing in the brain -- assuming that it follows similar computational principles.
Second, it enables machine learning applications on fast, low-power neuromorphic architectures that are specifically developed to mimic biological neuro-synaptic dynamics.
In this manuscript, we discuss several ways to answer what we consider to be a key challenge for neuromorphic architectures with analog components: Is it possible to design spiking architectures and training methods that are amenable to neuromorphic implementation and remain functionally performant despite substrate-inherent imperfections?

More specifically, we review three different approaches \cite{petrovici2015fast,petrovici2016robustness,schmitt2016classification}.
The first two are based on recent insights about how networks of spiking neurons can be constructed to sample from predefined joint probability distributions \cite{buesing2011neural,petrovici2016stochastic}.
When these distributions are learned from data, these networks automatically build an internal, generative model, which is then straightforward to use for pattern recognition and memory recall \cite{leng2016spiking}.
Practical problems arise when the hardware dynamics and parameter ranges are incompatible to the target specifications of the network, as these inevitably distort the sampled distribution.
The first approach involves the addition of auxiliary network components in order to make it robust to hardware-induced distortions (Sec.\,\ref{sec:sampling}).
The second one restricts the network topology in a way that endows it with immunity to some of these effects (Sec.\,\ref{sec:hierarchical}).
We demonstrate the effectiveness of both these approaches on the Spikey neuromorphic system \cite{pfeil2013six}.

The third strategy maps traditional feedforward architectures, trained offline with a backpropagation algorithm, to a network of spiking neurons on the neuromorphic device (Sec.\,\ref{sec:itl}).
Here, the key to good performance is an additional learning phase where parameters are trained on hardware in the loop, while using the abstract network description as an approximation for the parameter updates.
We show how this approach can restore network functionality despite having incomplete knowledge about the gradient along which the parameters need to descend.
These experiments are performed on the BrainScaleS neuromorphic system \cite{schemmel2010waferscale}.

While our networks are small compared to those used in contemporary machine learning applications, they showcase the potential of using accelerated analog neuromorphic systems for pattern representation and recognition.
In particular, the used neuromorphic systems operate $10^4$ times faster than their biological archetypes, thereby significantly speeding up both training and practical application.

\section{Fast sampling with spikes}
\label{sec:sampling}

\begin{figure}
    \centering
    \begin{tikzpicture}[]
        \draw[use as bounding box,inner sep=0pt] node {\includegraphics[width=\columnwidth]{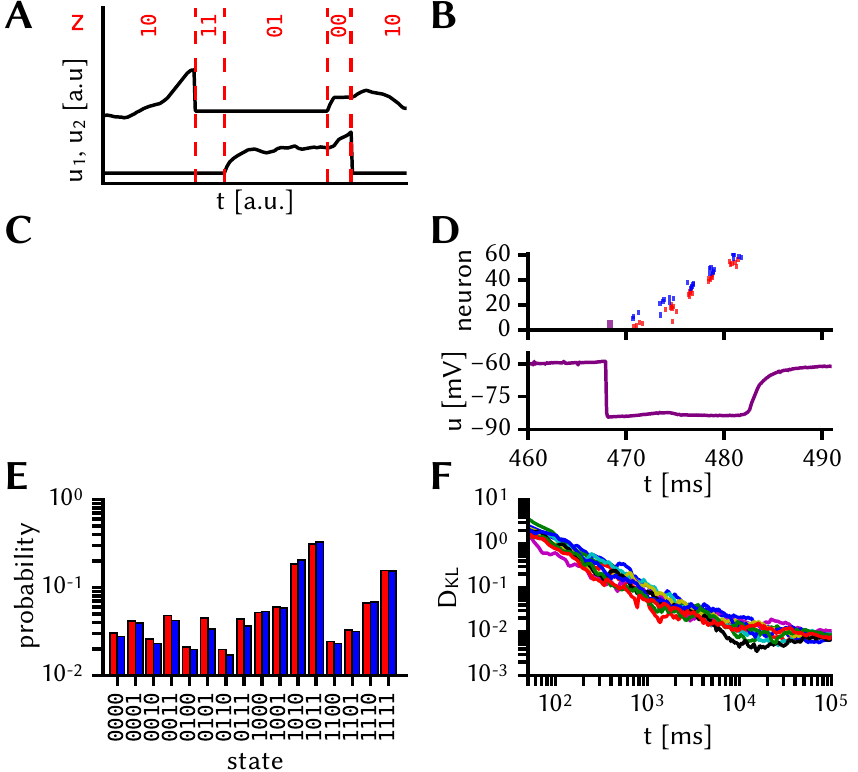}};
        \node at (-2.1, .5) {\includegraphics[width=.47\columnwidth]{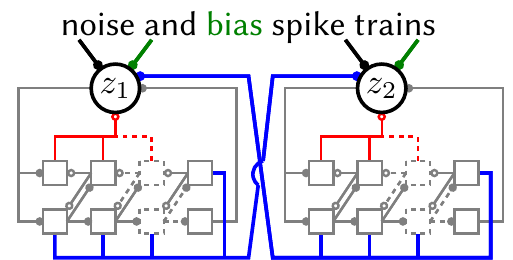}};
        \begin{scope}[
            line width=1pt,
            shift={(2.2,2.42)},
            font={\scriptsize \sffamily},
            ->,
            shorten >=1pt,
            shorten <=1pt,
            >=latex,
            ]
            \def \myop {.5}
            \tikzstyle{neuron}=[circle, draw=blue, inner sep=0pt, minimum size=12pt]
            \tikzstyle{hid}=[dashed, opacity=\myop]
            \tikzstyle{conn}=[-{>[flex=0.75]}, blue]
            \def \radius {.8cm}
            \def \arcmargin {17}
            \node[neuron] (z2) at (-30:\radius) {$z_2$};
            \node[neuron, hid] (z0) at (90:\radius) {};
            \node[neuron] (z1) at (210:\radius) {$z_1$};
            \foreach \i in {1,...,3}
            {
                \node (h\i) at (120*\i - 150:2.2*\radius) {};
            }
            \draw[conn] (z2) -- node[above,yshift=-2pt,xshift=2pt] {$w_{12}$} node[below,yshift=-9pt,xshift=-1pt] {$w_{21}$}(z1);
            \draw[conn, hid] (-30+\arcmargin:\radius) arc (-30+\arcmargin:90-\arcmargin:\radius);
            \draw[conn, hid] (z1) -- (z0);
            \draw[conn, hid] (90+\arcmargin:\radius) arc (90+\arcmargin:210-\arcmargin:\radius);
            \draw[conn, hid] (z0) -- (z2);
            \draw[conn] (210+\arcmargin:\radius) arc (210+\arcmargin:330-\arcmargin:\radius) ;
            \draw[black!40!green, hid, shorten <=5pt] (h2) to (z0);
            \draw[black!40!green] (h1) to node[above,shift=({5pt,-4pt})] {$b_2$} (z2);
            \draw[black!40!green] (h3) to node[above,shift=({-5pt,-4pt})] {$b_1$} (z1);
        \end{scope}
    \end{tikzpicture}
    \caption{Sampling with LIF neurons.
        \tb{(A)} Exemplary membrane potential traces and mapping of refractory/non-refractory neuron states to states 1/0 of binary RVs.
        \tb{(B)} Exemplary structure of a BM.
                 A subset of 2 units $(z_1, z_2)$ with biases $(b_1, b_2)$ (green) and connected by weights $w_{12}=w_{21}$ (blue) is highlighted to exemplify the neuromorphic network structure in subplot C.
        \tb{(C)} Sketch of sampling subnetworks representing binary RVs.
                 Each subnetwork consists of a principal LIF neuron (black circle) and an associated synfire chain that implements refractoriness (red synapses), and coupling between sampling units (blue synapses).
        \tb{(D)} Exemplary spike activity of a sampling unit and membrane potential of its PN.
        \tb{(E)} Target (blue) vs. sampled (red) distribution on the Spikey chip.
        \tb{(F)} Evolution of the Kullback-Leibler divergence between the sampled and the target distribution for multiple experimental runs.
                 Time given in biological units.
        }
    \label{fig:1}
\end{figure}
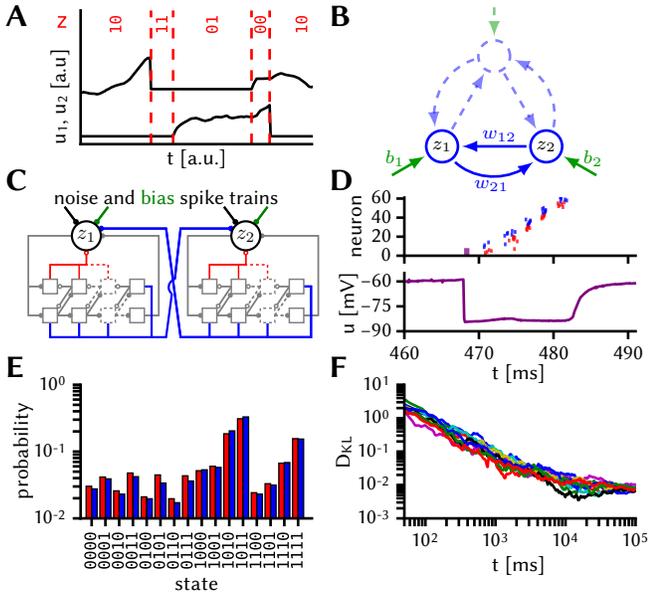

Following \cite{buesing2011neural,petrovici2016stochastic}, neural network activity can be interpreted as sampling from an underlying probability distribution over binary random variables (RVs).
The mapping from spikes to states $\bs z = (z_1,\dots,z_k)$ is defined by
\begin{equation}
    z_k^{(t)} = \left\{\begin{array}{ll}
                        1 \quad & \quad \text{if $t_k^s < t < t_k^s + \tauref$} \ , \\
                        0 \quad & \quad \text{otherwise} \ ,
                    \end{array}\right.
    \label{eqn:refractoriness}
\end{equation}
where $t_k^s$ are spike times of the $k$th neuron and $\tauref$ its absolute refractory period (Fig.\,\ref{fig:1}\,A).
When using leaky integrate-and-fire (LIF) neurons, Poisson background noise is used to achieve a high-conductance state, in which the stochastic response of a single neuron is well approximated by a logistic activation function
\begin{equation}
    p(z_k = 1) = \sigma \left([\bar u_k - \bar u_k^0]/\alpha\right) \ ,
    \label{eqn:actfctlif}
\end{equation}
where $\sigma(\cdot)$ is the logistic function and $\bar u_k$ represents the noise-free membrane potential of the $k$th neuron.
The parameters $\bar u_k^0$ (bias parameter determining the inflection point) and $\alpha$ (slope) are controlled by the intensity of the background noise.
With appropriate settings of synaptic weights $w_{ij}$ and bias parameters $\bar u_k^0$, these networks can be trained to sample from Boltzmann distributions
\begin{equation}
    p(\bs z) \propto \exp[-E(\bs{z})] = \exp \left[ \bs{z}^T \bs{W} \bs{z} / 2 + \bs{z}^T \bs{b} \right] \ ,
    \label{eqn:jointboltzmann}
\end{equation}
where the weight matrix $\bs{W}$ and the bias vector $\bs{b}$ can be chosen freely.
This enables the emulation of Boltzmann machines (BMs) with networks of LIF neurons (Fig.\,\ref{fig:1}\,B).

A core assumption of the neural sampling framework is that the membrane potential $u_k$ of a neuron reflects the state $\bs z_\nonk$ of all presynaptic neurons at any moment in time:
\begin{equation}
    u_k (\bs z_\nonk) = \textstyle\sum_{j \neq k}^n W_{kj} z_j + b_k \ .
    \label{eqn:uabstract}
\end{equation}
In particular, this requires that all neurons instantaneously transmit their states (spikes) to all their postsynaptic partners.
In any physical system, this assumption is necessarily violated to some degree, since signal transmission can never be instantaneous.
In the particular case of accelerated neuromorphic hardware, synaptic transmission delays become even more problematic, as they can be in the same order of magnitude as the state-encoding refractory times themselves.
Furthermore, the required equivalence between post-synaptic potential (PSP) durations and refractory states (\ref{eqn:refractoriness},\ref{eqn:uabstract}) can be violated if either of these are unstable.
On Spikey, for example, refractory times have relative spike-to-spike variations $\sigma_{\tauref}/\tauref$ between \SI{2}{\percent} and \SI{20}{\percent}.
These two kinds of timing mismatch pose a fundamental problem to the implementation of spiking BMs in accelerated analog substrates.

Here, we alleviate the issue of substrate-induced timing mismatches by using a recurrent network structure that represents each RV with a small subnetwork, called a sampling unit.
The subnetworks are built such that refractory times can be well controlled and, in addition, intra-unit refractory states and inter-unit state communication across the network are inseparably coupled (Fig.\,\ref{fig:1}\,C).

\begin{figure*}
\centering
        \begin{tikzpicture}
        \draw[use as bounding box,inner sep=0pt,anchor=south west] node {\includegraphics{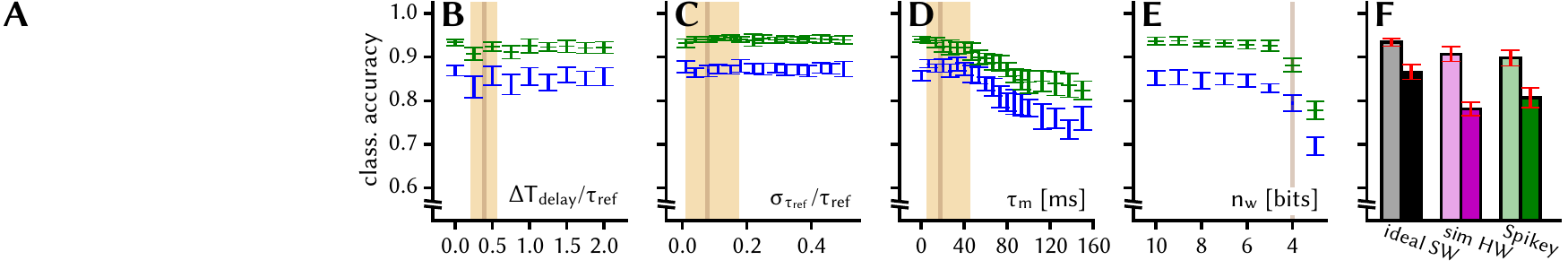}};
        \begin{scope}[
            shift={(0.1cm,0.1cm)},
            font={\scriptsize \sffamily},
            ->,
            shorten >=2pt,
            shorten <=2pt,
            >=latex,
            anchor=south west
            ]
            \def\layersep{1.225cm}
            \def\neuronsep{.9}
            \definecolor{viscol}{HTML}{008000}
            \colorlet{hidcol}{orange!75}
            \colorlet{labcol}{blue!75}
            \tikzstyle{neuron}=[circle,minimum size=15pt,inner sep=0pt]
            \tikzstyle{visible neuron}=[neuron, fill=viscol]
            \tikzstyle{hidden neuron}=[neuron, fill=hidcol]
            \tikzstyle{label neuron}=[neuron, fill=labcol]
            \def\numlab{2}
            \def\numhid{3}
            \def\numvis{4}
            \pgfmathsetmacro{\max}{max(\numvis, \numhid, \numlab)}
            \pgfmathsetmacro{\texthoroffset}{1sp}

            \foreach \x in {1,...,\numvis}
            \pgfmathparse{((\max - \numvis)*0.5 + \x - 1)*\neuronsep}
            \node[visible neuron] (V\x) at (\pgfmathresult,0) {};

            \foreach \x in {1,...,\numhid}
            {
                \pgfmathparse{((\max - \numhid)*0.5 + \x - 1)*\neuronsep}
                \node[hidden neuron] (H\x) at (\pgfmathresult, \layersep) {};
            }

            \foreach \x in {1,...,\numlab}
            {
                \pgfmathparse{((\max - \numlab)*0.5 + \x - 1)*\neuronsep}
                \node[label neuron] (L\x) at (\pgfmathresult, \layersep*2) {};
            }

            \foreach \source in {1,...,\numvis}
                \foreach \dest in {1,...,\numhid}
                    \draw[<->] (V\source) -- (H\dest);

            \foreach \source in {1,...,\numhid}
                \foreach \dest in {1,...,\numlab}
                    \draw[<->] (H\source) -- (L\dest);

            \node[anchor=west,right=\texthoroffset of V\numvis, align=center, text=viscol] (vl) {visible\\(144)};
            \node[anchor=west,right=\texthoroffset of H\numhid, align=center,text=hidcol] {hidden\\(50)};
            \node[anchor=west,right=\texthoroffset of L\numlab, align=center,text=labcol] {label\\(6)};
        \end{scope}
    \end{tikzpicture}
        \caption{
          Robustness from structure in hierarchical networks.
        \tb{(A)} Hierarchical spiking network emulating an RBM.
        \mbox{\tb{(B)--(E)}} Effects of hardware-induced distortions on the classification rate of the network.
                             Each test image was presented for a duration of \SI{1000}{\milli\second}.
                             Green: training data, blue: test data, brown: mean value and range of distortions measured on Spikey.
                             Error bars represent trial-to-trial variations.
        \tb{(B)} Synaptic transmission delays.
        \tb{(C)} Spike-to-spike variability of refractory times.
        \tb{(D)} Membrane time constant.
        \tb{(E)} Synaptic weight discretization.
        \tb{(F)} Comparison of classification rates in three scenarios: software simulation of the ideal, distortion-free case (black), software simulation of combined hardware-induced distortions as measured on Spikey (purple), hybrid emulation with the hidden layer on Spikey (green).
            Light colors for training data, dark colors for test data.
        }
    \label{fig:2}
\end{figure*}
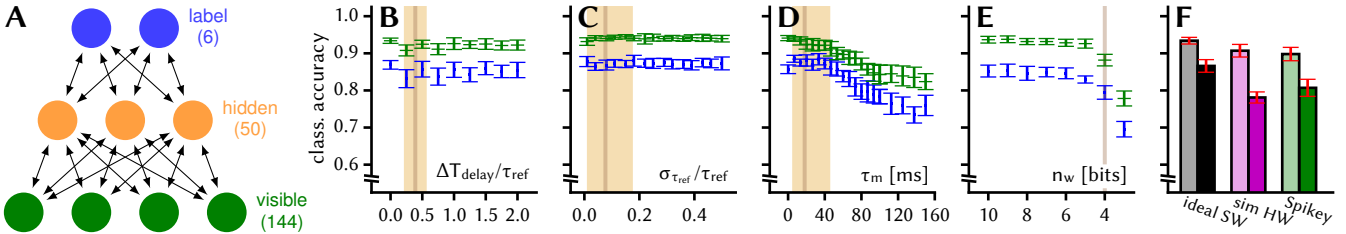

Sampling units consist of a single principle neuron (PN) and a small synfire chain of excitatory (EPs) and inhibitory populations (IPs). 
The EPs of each stage project to both populations in the following stage, thereby relaying an activity pulse in the forward direction.
The IPs project backwards, ensuring that neurons from previous stages only spike once.
Additionally, all IPs and the last EP also project onto the PN with large weights.
Therefore, after the PN elicits a spike, the IPs sequentially pull its membrane potential close to the inhibitory reversal potential, prohibiting it from firing as long as the synfire chain is active (Fig.\,\ref{fig:1}\,D).
When the pulse has reached the final synfire stage, its EP pulls the PN's membrane potential back to its equilibrium value.
The total duration of this pseudo-refractory period can then be controlled by the synfire chain length and parameters.

In addition to controlling refractoriness, the synfire chains also mediate the interaction between PNs.
The connections from a synfire chain to other PNs simply mirror its connections to its own PN.
This guarantees a match between effective interaction durations and pseudo-refractory periods.
The correct synapse parameter settings (weights, time constants) are determined in an iterative training procedure \cite{petrovici2015fast}.

The results of a hardware emulation can be seen in Fig.\,\ref{fig:1}\,E,\,F.
A network of four sampling units was trained on Spikey to sample from a target Boltzmann distribution.
After training, the network needs about \SI{e4}{\milli\second} of biological time to achieve a good match between the sampled and the target distribution.
Considering the hardware acceleration factor of $10^4$, this happens in \SI{1}{\milli\second} of wall-clock time.

\section{Robust hierarchical networks}
\label{sec:hierarchical}

As discussed in the previous section, sampling LIF networks are ostensibly sensitive to different types of hardware-induced timing mismatch.
In this subsection, we discuss how a sampling network model can be made robust by imposing a hierarchy onto the network structure \cite{petrovici2016robustness}.
This is the equivalent of moving from general BMs to restricted BMs (RBMs).
In addition to making their operation more robust, as we discuss below, this hierarchization has the distinct advantage of significantly speeding up training.

To emulate an RBM, we construct a hierarchical LIF network model with 3 layers: a visible layer representing the data, a hidden layer that learns particular motifs in the data and a label layer for classification (Fig.\,\ref{fig:2}\,A).
The network was trained with a contrastive learning rule
\begin{align}
    \Delta W_{ij} &\propto {\expect{z_i z_j}}_\mathrm{data} - \expect{z_i z_j}_\mathrm{model} \ , \label{eqn:contrastivew} \\
    \Delta b_i &\propto \expect{z_i}_\mathrm{data} - \expect{z_i}_\mathrm{model} \label{eqn:contrastiveb}
\end{align}
on a modified subset of the MNIST dataset ($\expect{\cdot}_\mathrm{data}$ and $\expect{\cdot}_\mathrm{model}$ represent expectation values when clamping training data and when the network samples freely, respectively).
Due to hardware limitations, we used a small network and dataset (6 digits, 12$\times$12 pixels, each with 20 training and 20 test samples) for this proof-of-principle experiment.

The specific influence of various hardware-induced distortion mechanisms were first studied in complementary software simulations.
These simulations show that the classification accuracy of the network is essentially unaffected by the types of timing mismatch discussed above, even when their amplitudes are much larger than those measured on our neuromorphic substrate (Fig.\,\ref{fig:2}\,B,\,C).
In order to facilitate a meaningful comparison with hardware experiments, two further distortion mechanisms were studied.
An upper limit to the membrane conductance can prevent neurons from entering a high-conductance state, thereby distorting their activation functions away from their ideal logistic shape (\ref{eqn:actfctlif}) and consequently modifying the sampled distribution.
However, within the range achievable on Spikey, the effect on the classification accuracy remains small (Fig.\,\ref{fig:2}\,D).
The largest effect (about \SI{5.6}{\percent} regression in classification accuracy compared to ideal software simulations) stems from the discretization of synaptic weights, which have a resolution of 4 bits on Spikey (Fig.\,\ref{fig:2}\,E).

The robustness of this hierarchical architecture to timing mismatches is a consequence of both the training procedure and the information flow within the network.
Training has the effect of creating a steep energy landscape $E(\bs z)$ (\ref{eqn:jointboltzmann}), for which deep energy minima, corresponding to particular learned digits, represent strong attractors, in which the system is placed during classification by clamping of the visible layer.
Throughout the duration of such an attractor, visible neurons represent pixels of constant intensity encoded in their spiking probability, thereby entering a quasi-rate-based information representation regime.
Therefore, the information they provide to the hidden layer is unaffected by temporal shifts or zero-mean noise.
As they outnumber the hidden neurons 24:1, they effectively control the state of the hidden layer.
The hidden layer neurons themselves are unaffected by timing mismatches because they are not interconnected.
Second-order (hidden$\rightarrow$label$\rightarrow$hidden) lateral interactions are indeed distorted, but as they are mediated by only few label neurons, their relative strength is too weak to play a critical role.

These findings are corroborated by experiments on Spikey (Fig.\,\ref{fig:2}\,F).
Due to the system's limitations, we used a hybrid approach, with the visible and label layers implemented in software and the hidden layer running on Spikey.
In the ideal, undistorted case, the LIF network had a classification performance of \SI{86.6 +- 1.7}{\percent} (\SI{93.4 +- 0.9}{\percent}) on the test (training) set.
This was reduced to \SI{78.1 +- 1.5}{\percent} (\SI{90.7 +- 1.7}{\percent}) when all distortive effects were simultaneously present in software simulations.
In comparison, the hybrid emulation showed a performance of \SI{80.7 +- 2.3}{\percent} (\SI{89.8 +- 1.8}{\percent}), which closely matched the software results within the trial-to-trial variability.
We stress that this was a result of direct-to-hardware mapping, with no additional training to compensate for hardware-induced distortions (as compared to Sec.\,\ref{sec:itl}).

\section{In-the-loop training}
\label{sec:itl}

\begin{figure}
  \centering
    \begin{tikzpicture}
    \draw[use as bounding box,inner sep=0pt] node {\includegraphics[width=\columnwidth]{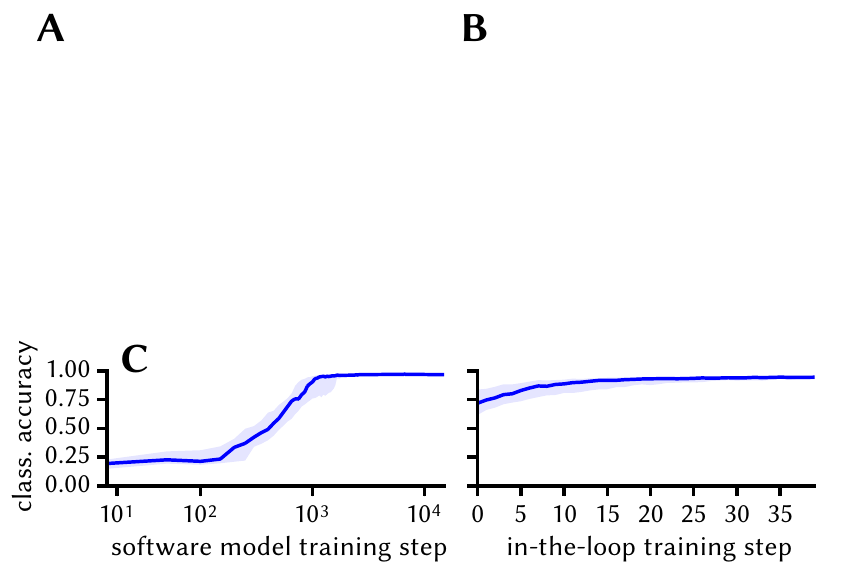}};
    \begin{scope}[
      shift={(-1.5in,-0.2cm)},
      font={\scriptsize \sffamily},
      ->,
      shorten >=2pt,
      shorten <=2pt,
      >=latex
      ]
            \def\layersep{0.9cm}
            \def\neuronsep{.9}
            \definecolor{viscol}{HTML}{008000}
            \colorlet{hidcol}{orange!75}
            \colorlet{labcol}{blue!75}
            \tikzstyle{neuron}=[circle,minimum size=13pt,inner sep=0pt]
            \tikzstyle{visible neuron}=[neuron, fill=viscol]
            \tikzstyle{hidden neuron}=[neuron, fill=hidcol]
            \tikzstyle{label neuron}=[neuron, fill=labcol]
            \def\numlab{2}
            \def\numhid{3}
            \def\numvis{4}
            \pgfmathsetmacro{\max}{max(\numvis, \numhid, \numlab)}
            \pgfmathsetmacro{\texthoroffset}{1sp}

            \foreach \x in {1,...,\numvis}
            \pgfmathparse{((\max - \numvis)*0.5 + \x - 1)*\neuronsep}
            \node[visible neuron] (V\x) at (\pgfmathresult,0) {};

            \foreach \x in {1,...,\numhid}
            {
                \pgfmathparse{((\max - \numhid)*0.5 + \x - 1)*\neuronsep}
                \node[hidden neuron] (H1\x) at (\pgfmathresult, \layersep) {};
            }

            \foreach \x in {1,...,\numhid}
            {
                \pgfmathparse{((\max - \numhid)*0.5 + \x - 1)*\neuronsep}
                \node[hidden neuron] (H2\x) at (\pgfmathresult, \layersep*2) {};
            }

            \foreach \x in {1,...,\numlab}
            {
                \pgfmathparse{((\max - \numlab)*0.5 + \x - 1)*\neuronsep}
                \node[label neuron] (L\x) at (\pgfmathresult, \layersep*3) {};
            }

            \foreach \source in {1,...,\numvis}
                \foreach \dest in {1,...,\numhid}
                    \draw[->] (V\source) -- (H1\dest);

            \foreach \source in {1,...,\numhid}
                \foreach \dest in {1,...,\numhid}
                    \draw[->] (H1\source) -- (H2\dest);

            \foreach \source in {1,...,\numhid}
                \foreach \dest in {1,...,\numlab}
                    \draw[->] (H2\source) -- (L\dest);

            \node[anchor=west,right=\texthoroffset of V\numvis, align=center, text=viscol] (vl) {visible\\(100)};
            \node[anchor=west,right=\texthoroffset of H1\numhid,
            align=center,text=hidcol] {hidden\\(15)};
            \node[anchor=west,right=\texthoroffset of H2\numhid, align=center,text=hidcol] {hidden\\(15)};
            \node[anchor=west,right=\texthoroffset of L\numlab, align=center,text=labcol] {label\\(5)};
        \end{scope}

    \begin{scope}[
            font={\scriptsize \sffamily},
            line width=2pt,
            -{>[flex=0.75]},
            >=latex,
            shift={(2.cm,1.cm)},
            line width=1pt
        ]
        \def \radius {1.2cm}
        \pgfmathsetmacro{\aangle}{acos(1.0/3.0)}
        \def \angleoffset {90} %use to rotate
        \def\data{{{"backpropagation",0,28,8},{"weight updates",\aangle,8,8},{"4\,bit weight discretization",180-\aangle,8,30},{"BrainScaleS",180,30,8},{"spikes",180+\aangle,8,8},{"ANN activity",360-\aangle,10,28}}}
        \pgfmathsetmacro{\numdata}{dim(\data)}

        \foreach \s [count=\i from 0] in {1,...,\numdata}
        {
            \pgfmathsetmacro{\c}{\data[\i][1] + \angleoffset}
            \node[fill=white,inner sep=1pt] (\s) at (\c:\radius) {\pgfmathparse{\data[\i][0]}\pgfmathresult};
            \pgfmathsetmacro{\arcstart}{mod(\c+\data[\i][2],360)}
            \pgfmathsetmacro{\arcend}{\data[mod(\s,\numdata)][1]+\angleoffset-\data[\i][3]}
            \draw (\arcstart:\radius) arc (\arcstart:\arcend:\radius);
        }
        \def \hordist {.5cm}
        \node[anchor=base] (7) [left=\hordist of 4]  {MNIST};
        \node[anchor=base] (8) [right=\hordist of 4]  {prediction};
        \draw (7) -- (4);
        \draw (4) -- (8);
        \node (9) [below=-2pt of 7] {};
        \draw[color=purple!75] (9) --  (9-|8.south);
        \node[below=.1cm of 4,color=purple!75] {forward pass};
        \node[above=3pt of 1,color=purple!75] {backward pass};
        \begin{scope}[on background layer]
            \pgfmathsetmacro{\arcstart}{-35}
            \pgfmathsetmacro{\arcend}{215}
            \def\radiusmult {1.2}
            \draw[line width=2pt,color=purple!75] (\arcstart:\radiusmult*\radius) arc (\arcstart:\arcend:\radiusmult*\radius);
        \end{scope}
    \end{scope}
  \end{tikzpicture}
  \caption{In-the-loop training.
        \tb{(A)} Structure of the feed-forward, rate-based deep spiking network.
        \tb{(B)} Schematic of the training procedure with the hardware in the loop.
        \tb{(C)} Classification accuracy over training step.
                 Left: software training phase, right: hardware in-the-loop training phase.
        }
    \label{fig:3}
\end{figure}
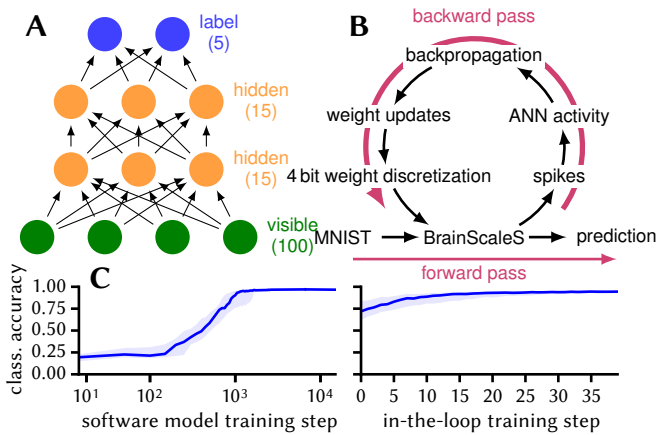

In Sec.\,\ref{sec:sampling}, we used a training procedure based on (\ref{eqn:contrastivew},\ref{eqn:contrastiveb}) to optimize the hardware-emulated sampling network.
Such simple contrastive learning rules can yield very good classification performance in networks of spiking neurons \cite{leng2016spiking}.
Another class of highly successful learning algorithms is based on error backpropagation.
This, however, requires precise knowledge of the gradient of a cost function with respect to the network parameters, which is difficult to achieve on analog hardware.
We propose a training method for hardware-emulated networks that circumvents this problem by using the cost function gradient with respect to the parameters of an ANN as an approximation of the true gradient with respect to the hardware parameters \cite{schmitt2016classification}.
A similar method has previously been used for network training on a digital neuromorphic device \cite{esser2016convolutional}.

Our training schedule consisted of two phases.
In the first phase, an ANN was trained in software on a modified subset of the MNIST dataset (5 digits, 10$\times$10 pixels, with a total of 30690 training and 5083 test samples) using a simple cost function with regularization
\begin{equation}
    \textstyle C(\bs W) = \sum_{s \in S} \left\Vert \bs{\tilde{y}}_{s} - \bs{\hat{y}}_{s}) \right\Vert ^2 + \sum_{kl} \tfrac{1}{2}\lambda W_{kl}^2
    \label{eq:cost}
\end{equation}
and backpropagation with momentum \cite{qian1999momentum}
\begin{align}
    \Delta W_{kl} &\leftarrow \eta \nabla_{W_{kl}} C(\bs W) + \gamma \Delta W_{kl} \ , \\
    W_{kl} &\leftarrow W_{kl} - \Delta W_{kl} \ .
\end{align}
Here, $\bs{\tilde{y}}_{s}$ and $\bs{\hat{y}}_{s}$ denote the target and network state of the label layer, respectively, and the sum runs over all samples within a minibatch $S$.
The learned parameters were then translated to a feed-forward spiking neural network (Fig.\,\ref{fig:3}\,A).
Here, the BrainScaleS wafer-scale system \cite{schemmel2010waferscale} was used for network emulation.
Due to hardware imperfections, the ANN classification accuracy of \SI{97}{\percent} dropped to \asymunc{72}{12}{10}{\si{\percent}} after mapping the network to the hardware substrate.

In the second training phase, the hardware-emulated network was trained in the loop (Fig.\,\ref{fig:3}\,B) for several iterations.
Parameter updates were calculated using the same gradient descent rule as in the ANN, but the activation of all layers was measured on the hardware.
The rationale behind this approach is that the activation function of an ANN unit is sufficiently similar to that of an LIF neuron to allow using the computed gradient as an approximation of the true hardware gradient.
As seen in Fig.\,\ref{fig:3}\,C, this assumption is validated by the post-training performance of the hardware-emulated network: after 40 training iterations, the classification accuracy increased back to \asymunc{95}{1}{2}{\si{\percent}}.

\section{Discussion}

We have reviewed three strategies for emulating performant spiking network models in analog hardware.
The proposed methods tackled the problems induced by substrate-inherent imperfections from different (and complementary) angles.
The three strategies were implemented and evaluated with two different analog hardware systems.

An essential advantage of the employed neuromorphic platforms is provided by their accelerated dynamics.
Despite possible losses in performance compared to precisely tunable software solutions, accelerated analog neuromorphic systems have the potential to vastly outperform classical simulations of neural networks in terms of both speed and energy consumption \cite{schmitt2016classification} -- an invaluable advantage for on-line learning of complex, real world data sets.
The network in Sec.\,\ref{sec:sampling}, for example, is already faster than equivalent software simulations (NEST 2.2.2 default build, single-threaded, Intel Core i7-2620M) by several orders of magnitude.

The studied networks serve as a proof of principle and are scalable to larger network sizes.
Future research will have to address whether the results obtained for these small networks still hold as training tasks increase in complexity.
Furthermore, the generative properties of the described hierarchical LIF networks remain to be studied.
Another major step forward will be taken once training can take place entirely on the hardware, thereby rendering sequential reconfigurations between individual experiments unnecessary.
Future generations of the used systems will feature on-board plasticity processor units, with early-stage experiments already showing promising results \cite{friedmann2016demonstrating}.

\section*{Acknowledgments}

The first five authors contributed equally to this work.
This research was supported by EU grants \#269921 (BrainScaleS), \#604102 and \#720270 (Human Brain Project) and the Manfred Stärk Foundation.

\bibliographystyle{IEEEtran}

\end{document}